\documentclass[twocolumn]{aastex62}

\newcommand{\Msun}{\,M_{\odot}}

\def\teff{T_{\rm eff}}
\def\msun{M_{\odot}}

\def\logg{{\rm log}\,g}
\def\loggast{{\rm log}\,g_{\rm astero}}

\def\amlt{\alpha_{\rm MLT}}

\def\tauphot{\tau_{\rm phot}}
\def\Ttau{T\textrm{--}\tau}

\graphicspath{{./}{figures/}}

\submitjournal{ApJ}

\shortauthors{CHOI ET AL.}
\shorttitle{On the Red Giant Branch}

\begin{document}

\title{ON THE RED GIANT BRANCH: AMBIGUITY IN THE SURFACE BOUNDARY CONDITION LEADS TO $\approx100$~K UNCERTAINTY IN MODEL EFFECTIVE TEMPERATURES}

\author[0000-0002-8822-1355]{Jieun Choi}
\affiliation{Harvard--Smithsonian Center for Astrophysics, 60 Garden Street, Cambridge, MA 02138, USA}
\author[0000-0002-4442-5700]{Aaron Dotter}
\affiliation{Harvard--Smithsonian Center for Astrophysics, 60 Garden Street, Cambridge, MA 02138, USA}
\author[0000-0002-1590-8551]{Charlie Conroy}
\affiliation{Harvard--Smithsonian Center for Astrophysics, 60 Garden Street, Cambridge, MA 02138, USA}
\author[0000-0001-5082-9536]{Yuan-Sen Ting}
\affiliation{Institute for Advanced Study, Princeton, NJ 08540, USA}
\affiliation{Department of Astrophysical Sciences, Princeton University, Princeton, NJ 08544, USA}
\affiliation{Observatories of the Carnegie Institution of Washington, 813 Santa Barbara Street, Pasadena, CA 91101, USA}

\begin{abstract}
The effective temperature ($\teff$) distribution of stellar evolution models along the red giant branch (RGB) is sensitive to a number of parameters including the overall metallicity, elemental abundance patterns, the efficiency of convection, and the treatment of the surface boundary condition. Recently there has been interest in using observational estimates of the RGB $\teff$ to place constraints on the mixing length parameter, $\amlt$, and possible variation with metallicity. Here we use 1D \texttt{MESA} stellar evolution models to explore the sensitivity of the RGB $\teff$ to the treatment of the surface boundary condition. We find that different surface boundary conditions can lead to $\pm100$ K metallicity-dependent offsets on the RGB relative to one another in spite of the fact that all models can reproduce the properties of the Sun. Moreover, for a given atmosphere $\Ttau$ relation, we find that the RGB $\teff$ is also sensitive to the optical depth at which the surface boundary condition is applied in the stellar model. Nearly all models adopt the photosphere as the location of the surface boundary condition but this choice is somewhat arbitrary. We compare our models to stellar parameters derived from the APOGEE-{\it Kepler} sample of first ascent red giants and find that systematic uncertainties in the models due to treatment of the surface boundary condition place a limit of $\approx100$~K below which it is not possible to make firm conclusions regarding the fidelity of the current generation of stellar models.
\end{abstract}



\section{Introduction}
\label{s:intro}

Models of the red giant branch (RGB) underpin much of our knowledge of nearby resolved dwarf galaxies \citep[e.g.,][]{Tolstoy2009, Weisz2011}, the formation history of various components of the Milky Way \citep[e.g.,][]{Freeman2002, Rix2013}, photometric metallicities of halo stars in nearby galaxies \citep[e.g., PAndAS,][]{McConnachie2009, Harris2002}, and the stellar population properties of distant galaxies \citep[e.g.,][]{Walcher2011, Conroy2013}.

Despite a long history \citep[e.g.,][]{Hoyle1955, Demarque1963}, accurate modeling of the RGB remains a challenge. The effective temperature ($\teff$) of the RGB is sensitive to changes in chemical composition, stellar mass, and model parameters such as convection efficiency, and there are essentially no ``ground truth'' observations with which to constrain the models in this evolutionary phase. This is important because 1D stellar evolution models treat various processes including convection in a phenomenological way with one or more free parameters that must be calibrated in some way. The most common framework for convection, the so-called mixing length theory \citep[MLT;][]{BohmVitense1958}, utilizes a free parameter of order unity, $\amlt$, to describe the convection efficiency. Calibration of $\amlt$ in the current generation of stellar evolution models is primarily based upon exquisite knowledge of a single star --- the Sun. It is important to test whether the free parameters governing convection that have been calibrated to the Sun are capable of accurately modeling red giants as well.

There have been several lines of evidence, both from a theoretical perspective \citep[e.g.,][]{Trampedach2014, Magic2015} and from comparing models to observations \citep[e.g.,][]{Salaris1996, Bonaca2012, Tayar2017, Joyce2017, Chun2018, Li2018, Viani2018}, that suggest the properties of convection may vary with stellar parameters, such as $\logg$, $\teff$, and/or metallicity. If true, this would have profound implications extending well beyond the realm of stellar physics. For example, in the case of a solar metallicity, $\logg \approx 2$ star, increasing $\amlt$ by $\approx 0.1$ ($5~\%$) would make the RGB hotter by $\approx 50$~K, or equivalently $\approx 0.02$~mag in $B-V$ and $J-K$. This seemingly small change shifts the isochrone-based mass and age estimates upward by $\rm \approx 0.25~\msun$ and downward by a factor of $\approx 2$, respectively. This is an uncomfortable margin of error when they consider e.g., RGB-based star formation histories of the Milky Way and nearby resolved galaxies. 

Recently, \citet[][hereafter T17]{Tayar2017} used the joint APOGEE-\emph{Kepler} catalog of red giants \citep[][]{Pinsonneault2015} to compare observationally-derived $\teff$ values to YREC \citep{Pinsonneault1989} and PARSEC \citep{Bressan2012} stellar evolution models. T17 found that the model $\teff$ values were too hot and cool at low and high metallicities, respectively. They interpreted this metallicity-dependent discrepancy between the YREC and APOGEE $\teff$ as evidence for a metallicity-dependent convection efficiency parameter. \citet[][hereafter S18]{Salaris2018} revisited the T17 analysis with their own stellar models from the BaSTI collaboration \citep{Pietrinferni2004}. These authors also reported a discrepancy, albeit smaller, between their predicted and APOGEE $\teff$ for the full T17 sample. However, they found that this tension mostly disappears when considering a subsample of metal-rich, solar-scaled stars.

One of the key points from S18 is the importance of the surface boundary condition (BC) in setting the $\teff$ distribution of their model RGB. There have been several studies in the literature on the effect of BCs, both on the $\teff$ and the overall stellar structure \citep[e.g.,][]{Chabrier1997, Montalban2001, Salaris2002, VandenBerg2008} and on the lithium depletion boundary technique for age-dating clusters \citep{Burke2003}. Given the importance of this issue, we revisit the $\teff$ of the RGB in the context of \texttt{MESA} stellar evolution models. In Section~\ref{s:models}, we introduce the stellar models and discuss the use of different BCs in model computations. This is followed by Section~\ref{s:mvd} where we critically examine the $\teff$ discrepancy as a function of metallicity. We place these results in context in Section \ref{s:discussion}, and we provide a summary in Section~\ref{s:summary}.

\begin{figure*}[!t]
\center
\includegraphics[width=0.85\textwidth]{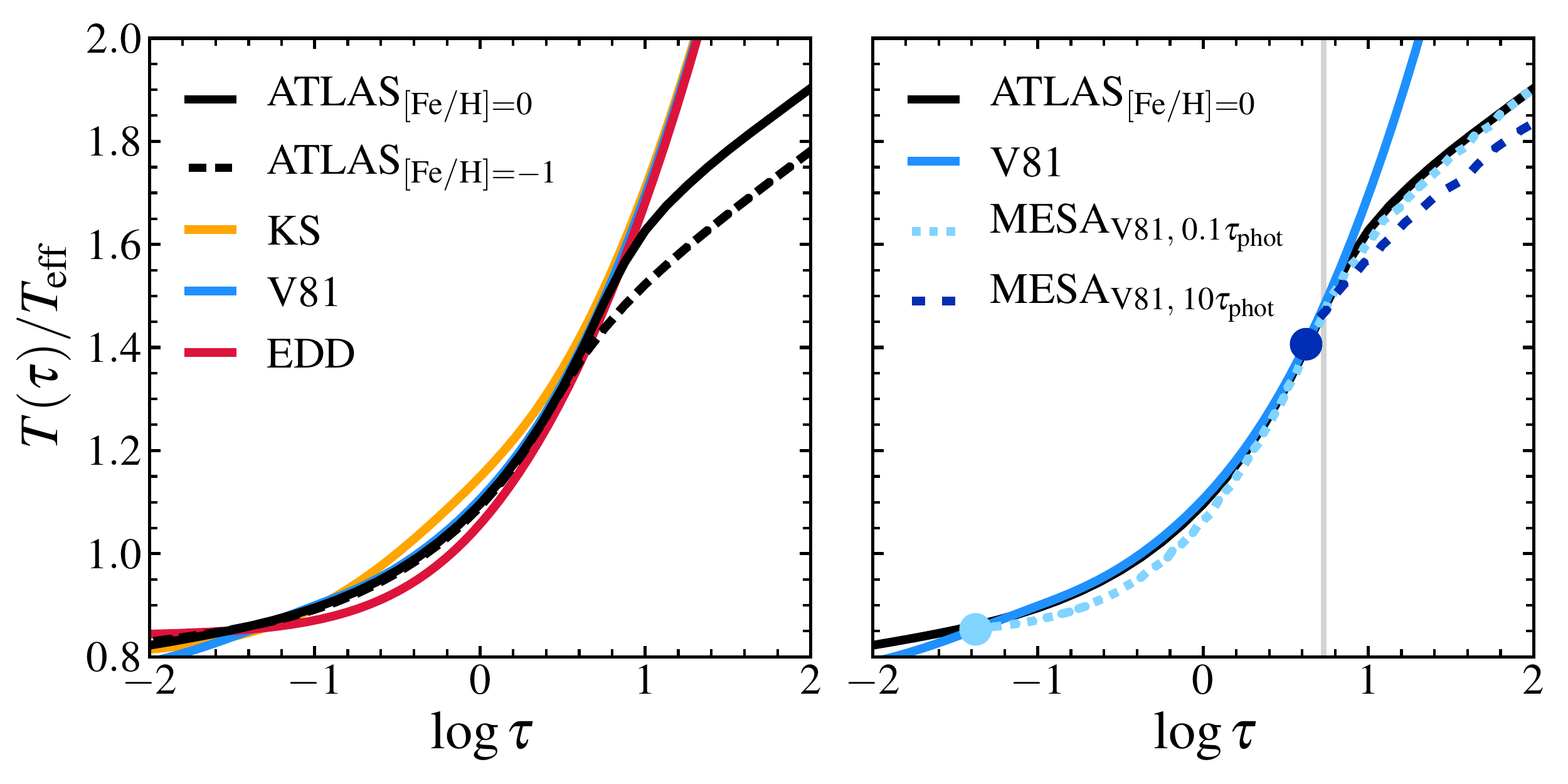}
\vspace{0.1cm}
\caption{Overview of the various atmosphere $\Ttau$ profiles considered in this work. Left Panel: The \cite{KrishnaSwamy1966}, \cite{Vernazza1981}, and \cite{Eddington1926} analytic relations are shown in orange, blue, and red lines, respectively. The \texttt{ATLAS} model atmospheres at $\logg=2$ and $\teff=4750$~K at both [Fe/H]=$-1$ and $0$ are plotted as black lines. The analytic $\Ttau$ relations increase sharply above $\log \tau \approx 1$ while the \texttt{ATLAS} model profiles become shallower due to the onset of convection in the latter. Right Panel: Comparison of the atmosphere $\Ttau$ profiles for \texttt{ATLAS} and V81 with the \texttt{MESA} stellar interior profiles computed using the V81 BC. Light and dark blue circles mark the outer edges of the \texttt{MESA} profiles, at optical depths of $0.1\tauphot$ ($\tau_{\rm base}=0.042$) and and $10\tauphot$ ($\tau_{\rm base}=4.2$), respectively. The \texttt{MESA} profiles become shallower at large optical depths due to the onset of convection. The gray vertical line marks the optical depth above which the convective flux is greater than the radiative flux.}
\label{fig:Ttau}
\end{figure*}


\section{Stellar Models}
\label{s:models}

\subsection{Input Physics}

All stellar evolutionary tracks used in this paper are computed using Modules for Experiments in Stellar Astrophysics \citep[\texttt{MESA};][]{Paxton2011, Paxton2013, Paxton2015, Paxton2018}, revision 9793. The physical assumptions and input data sources are consistent with the \texttt{MESA} Isochrones and Stellar Tracks \citep[][]{Dotter2016, Choi2016} project. To summarize, we use opacities from OPAL \citep{Iglesias1996} and AESOPUS \citep{Marigo2009}, and equation of state from OPAL \citep{Rogers2002}, HELM \citep{Timmes2000}, and SCvH \citep{Saumon1995} blended as described by \citet{Paxton2011}. We treat atomic diffusion using the formalism of \citet{Thoul1994} with turbulent mixing at the surface as described by \citet{Dotter2017}. Nuclear reaction rates are taken from JINA Reaclib v2.2 \citep{Cyburt2010}. Convective boundary mixing is treated in the diffusive approximation with the exponential decay formula \citep{Freytag1996} and a value of $f_{\rm ov}=0.016$. We adopt the \citet{Asplund2009} solar abundance pattern throughout.

\subsection{Surface Boundary Conditions}

The surface BC is an essential input to any stellar model as it is required to close the equations of stellar structure. Conceptually, this can be thought of as the point at which the stellar ``atmosphere'' is attached to the stellar interior. In practice the BC is set by specifying the pressure and temperature at the last grid point in the interior model, or the stellar ``surface." The surface is commonly defined as the point at which $T=\teff$; we will refer to this location as the photosphere (with a corresponding Rosseland optical depth of $\tauphot$). However, in some cases the BC is set much deeper in the atmosphere \citep[e.g., at $\tau=100$;][]{Chabrier1997, Choi2016}. A more important criterion is that the joining region should be located where the various assumptions and adopted microphysics are in agreement between the atmosphere and the interior.

Most stellar evolution codes approach the treatment of the surface BC in one of two ways: integration of analytic $\Ttau$ relations and model stellar atmosphere tables. We consider both approaches in this paper.

The classic means of obtaining the surface BCs for $T$ and $P$ is to adopt an analytic $\Ttau$ relation and integrate $dP/d\tau=(g/\kappa) - (a/3)dT^4/d\tau$ from $\tau \approx 0$ to $\tau_{\rm base}$, with $\tau_{\rm base} = \tauphot$ being the most common choice. Here, $g$ and $\kappa$ are the local surface gravity and Rosseland mean opacity, respectively, and the second term accounts for the radiation pressure where $a$ is the radiation constant. Once $\tau_{\rm base}$ is chosen, $T(\tau_{\rm base})$ is trivially obtained from the $\Ttau$ relation, and $P(\tau_{\rm base})$ is obtained from the above integral.

The analytic $\Ttau$ relations considered in this paper include: i) EDD: the \citet{Eddington1926} ``gray'' $\Ttau$ relation for which $\tauphot=2/3$; ii) KS: the fit by \citet{KrishnaSwamy1966} for which $\tauphot\approx 0.31$; and iii) V81: an analytical fit to the \citet{Vernazza1981} solar atmosphere from S18 for which $\tauphot\approx 0.42$. These $\Ttau$ relations are shown for a typical red giant star in Figure \ref{fig:Ttau}. An important (and relatively untested) assumption of analytic $\Ttau$ relations is that they do not depend on chemical composition (metallicity) nor on surface gravity. 

Another approach to the surface BC is to use model atmospheres (as in e.g., Lyon, \citealt{Baraffe2015}; MIST, \citealt{Choi2016}), which tabulate thermodynamic and other microphysical quantities as a function of optical depth, to determine the pressure at some reference location in the atmosphere. For the purposes of this paper we use model atmosphere BC tables constructed at $\tau=\tauphot$. The model atmosphere surface BC in this work is based on grids of \texttt{ATLAS12} \citep{Kurucz1970, Kurucz1993} models. The gas pressures at $\tauphot$ are tabulated as a function of $\teff$, $\logg$, and metallicity, and then interpolated by \texttt{MESA} at runtime (see \citealt{Paxton2011, Choi2016} for further details). 

\begin{figure*}[!t]
\center
\includegraphics[width=0.85\textwidth]{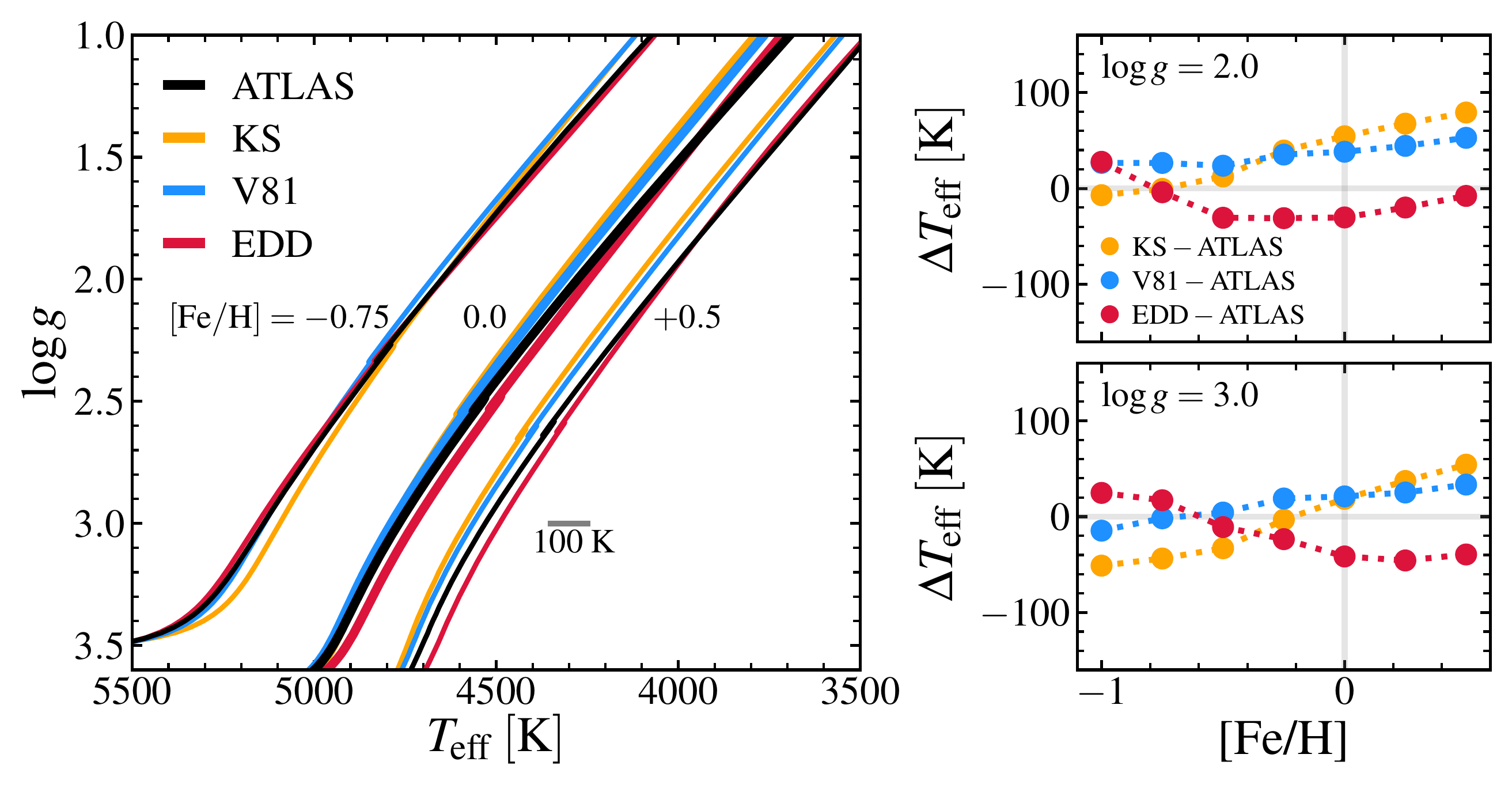}
\vspace{0.1cm}
\caption{Effect of varying boundary conditions in the \texttt{MESA} model calculations on the RGB effective temperatures. Left Panel: RGB sequences in the Kiel diagram for a $1.2~\msun$ star at [Fe/H]$=-0.75$, 0.0, and $+0.5$. The black curves corresponds to the fiducial \texttt{MESA} models computed with \texttt{ATLAS} photosphere tables. Each set of models for a given BC has been independently calibrated to the Sun. Right Panels: $\Delta \teff$ for the RGB models computed with the analytic $\Ttau$ models relative to the \texttt{ATLAS} RGB $\teff$ as a function of metallicity at two $\logg$ values. }
\label{fig:dtm_varyBCs}
\end{figure*}

\begin{figure*}
\center
\includegraphics[width=0.85\textwidth]{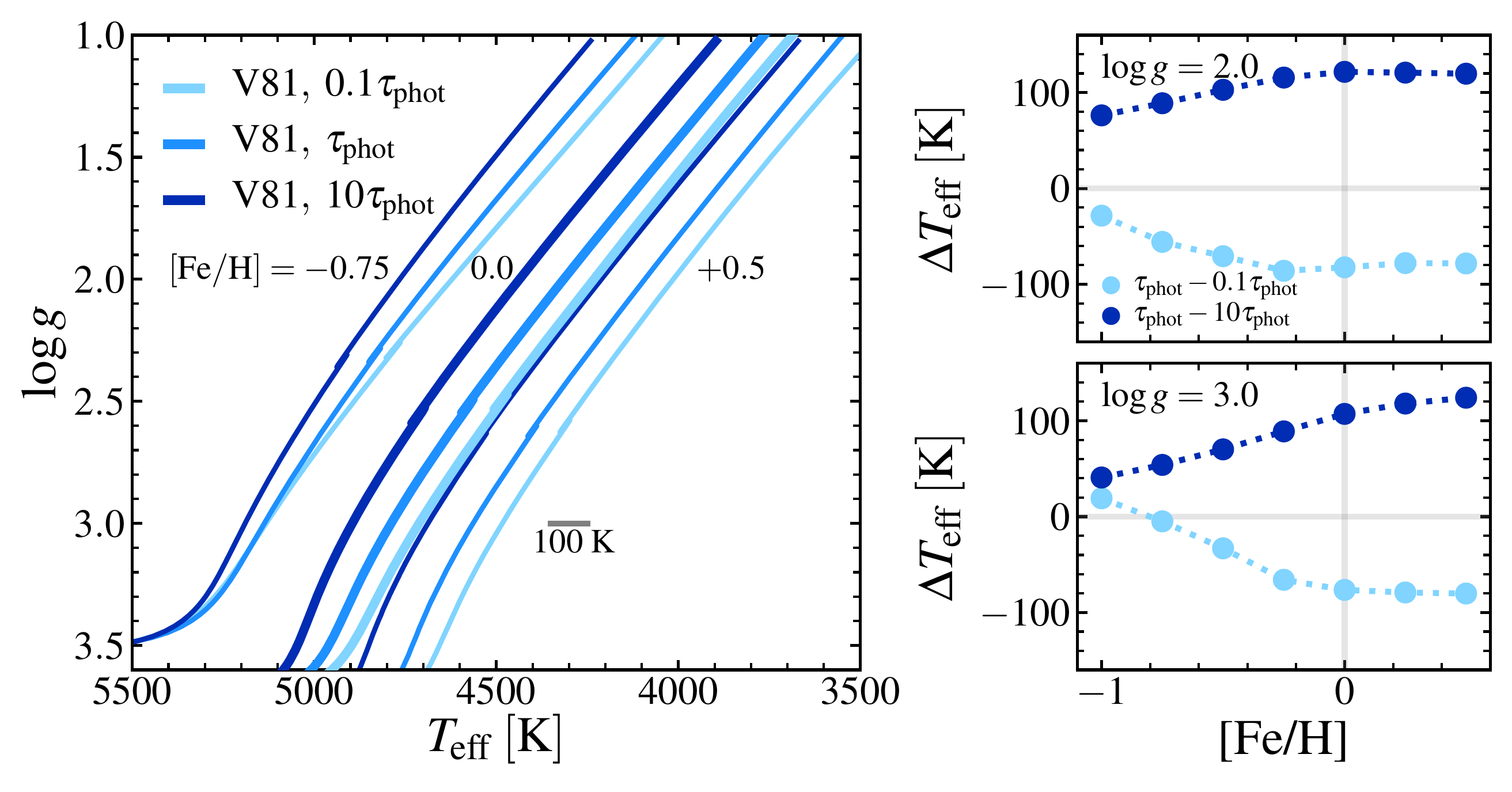}
\vspace{0.1cm}
\caption{Same as Figure~\ref{fig:dtm_varyBCs} now varying $\tau_{\rm base}$, the optical depth at which the V81 surface boundary condition is applied in the \texttt{MESA} model. Each set of models has been independently calibrated to the Sun. Depending on one's choice of $\tau_{\rm base}$, there is an approximately 100~K variation in the resulting RGB $\teff$ relative to the nominal $\tau_{\rm base}=\tauphot$ case over the plotted metallicity range.}
\label{fig:dtm_varyBCs_v2}
\end{figure*}

Figure \ref{fig:Ttau} shows the $\Ttau$ relations extracted from two \texttt{ATLAS} models along with the analytic versions. We show an \texttt{ATLAS} model with $\teff=4750$~K, $\logg=2$, and $\rm [Fe/H]=0$ for which $\tauphot \approx 0.42$ and another \texttt{ATLAS} model with $\teff=4750$~K, $\logg=2$, and $\rm [Fe/H]=-1$ for which $\tauphot \approx 0.44$. Interestingly, $\tauphot$ in both \texttt{ATLAS} models are very similar to the V81 relation. Indeed, the similarity with V81 extends from $-1 \leq \log\tau \leq 0.5$. The metallicity-dependence appears to be mild for $\tau<4$. 

In this paper we compute stellar evolutionary tracks for three analytic $\Ttau$ relations and for the \texttt{ATLAS} model atmosphere tables. Each surface BC is separately calibrated to reproduce the solar parameters at the solar age as in \citet{Choi2016}. The main parameter that changes as one considers different BCs is $\amlt$. Solar calibration as a means to obtain $\amlt$ is standard practice in almost all stellar models (see \citealt{Ferraro2006} where the authors calibrate $\amlt$ using globular clusters instead), but there is no guarantee that this is sufficient to accurately model stars in other evolutionary phases or at non-solar abundances and masses.

For the \texttt{ATLAS} BC tables we obtain a solar-calibrated $\amlt=1.848$; for the Eddington $\Ttau$ $\amlt=1.713$; for the V81 $\Ttau$ $\amlt=1.908$; and for the Krishna-Swamy $\Ttau$ $\amlt=2.008$. The solar-calibrated $\amlt$ values vary by $\approx15\%$ across the range of surface BCs considered here. As pointed out by S18, the solar-calibrated $\amlt$ values derived from the \citet{Vernazza1981} $\Ttau$ relation and model photosphere tables are closer to each other than to either of the other two $\Ttau$ relations. As illustrated in Figure~\ref{fig:Ttau}, these two BCs have very similar $\Ttau$ profiles. 

\subsection{Influence of the Surface Boundary Condition on the Effective Temperature of the RGB}

We compute grids of stellar evolutionary tracks for the four surface BCs and their solar-calibrated $\amlt$ values, for initial masses between 0.7 and 2.5~$\msun$ in steps of 0.1~$\msun$ and [Fe/H]$=-1$ to +0.5 in steps of 0.25~dex, and for [$\alpha$/Fe]=0, $+0.2$, and $+0.4$. We assume that all of the $\alpha$-capture elements (O, Ne, Mg, Si, S, Ar, Ca, and Ti) are enhanced by the same amount as denoted by [$\alpha$/Fe].

The left panel of Figure~\ref{fig:dtm_varyBCs} shows RGB tracks computed for the four surface BCs and their solar-calibrated $\amlt$ values in the Kiel diagram. The evolutionary tracks correspond to $M_{\rm init} = 1.2~\msun$ and [Fe/H]=$-0.75$, 0, $+0.5$, all with [$\alpha$/Fe]=0. The right panels show how $\teff$ varies at fixed $\logg$ over the full range of [Fe/H] relative to the fiducial (ATLAS BC) \texttt{MESA} models. The maximum difference in $\teff$ is about 100~K with the Eddington $\Ttau$ BC usually---but not always---producing the coolest RGB and the Krishna-Swamy $\Ttau$ BC usually producing the hottest RGB. Aside from these generalities, the $\Delta \teff$ behavior is complex, varying both as a function of [Fe/H] and $\logg$.
 
In order to explore the implications of changing $\tau_{\rm base}$ within the context of a given surface BC, we have computed a series of models with the V81 $\Ttau$ relation but now shifting the fitting point to $\tau_{\rm base}= 0.1\tauphot$ and $10\tauphot$, each with its own solar-calibrated $\amlt$. The V81 models with $\tau_{\rm base}=0.1\tauphot$ have $\amlt=1.71$ while those with $\tau_{\rm base}=10\tauphot$ have $\amlt=2.106$. We remind the reader that V81 and $\tau_{\rm base}=\tauphot$ has a solar-calibrated $\amlt=1.908$. We chose the V81 $\Ttau$ for this exercise both because it follows the \texttt{ATLAS} model $\Ttau$ closely and because \texttt{MESA} provides a convenient option to change the location of $\tau_{\rm base}$ for a given analytic $\Ttau$ relation. In contrast, the same test with \texttt{ATLAS} would require the calculation and implementation of an entirely new set of atmosphere tables. Nevertheless, given their similarities in Figure~\ref{fig:Ttau}, we believe the conclusions drawn from the V81 $\Ttau$ relation should be applicable to the \texttt{ATLAS} atmosphere tables. 

Figure~\ref{fig:dtm_varyBCs_v2} illustrates the effect of changing $\tau_{\rm base}$ on the location of the RGB in the Kiel diagram. The $\teff$ shifts are mainly due to differences in the solar-calibrated $\amlt$ for each assumed $\tau_{\rm base}$. Notice that the $\teff$ shifts are quite large, of order $\pm100$ K at fixed metallicity, with strong variation as a function of $\logg$ and [Fe/H]. This indicates that there is substantial model uncertainty in the effective temperature distribution along the RGB due solely to the ambiguity in where one sets the surface BC.


\newpage

\section{Models vs. Data}
\label{s:mvd}

\subsection{Observations on the RGB}
\label{s:data}

Now we turn our attention to the comparison between the model-predicted RGB $\teff$ with observational estimates of $\teff$ for real stars. We use the publicly available T17 catalog\footnote{\texttt{www.astronomy.ohio-state.edu/$\sim$tayar/MixingLength.htm}}, which contains spectroscopic $\teff$, [Fe/H], and [$\alpha$/Fe], as well as asteroseismic $\logg$ and $M$. As described in T17, there are approximately 12,000 APOGEE stars with {\it Kepler} asteroseismic observations (i.e., APOKASC), and of those, 3210 are identified as first ascent red giants. Measurements of two global asteroseimic parameters, $\Delta \nu$ and $\nu_{\rm max}$, can be combined with an external $\teff$ estimate to derive estimates for stellar mass and radius (or $\logg$): $M \propto \nu_{\rm max}^3 \,\Delta \nu^{-4} \,\teff^{3/2}$ and $g_{\rm astero} \propto \nu_{\rm max} \,\,\teff^{1/2}$ \citep{Kjeldsen1995},
where the proportionality constants are defined by scaling to the solar values. The $\teff$ values in T17 were corrected for a metallicity-dependent offset that was determined by a comparison to the color-$\teff$ relation from \citet{GonzalezHernandez2009}.

\begin{figure}
\center
\includegraphics[width=0.45\textwidth]{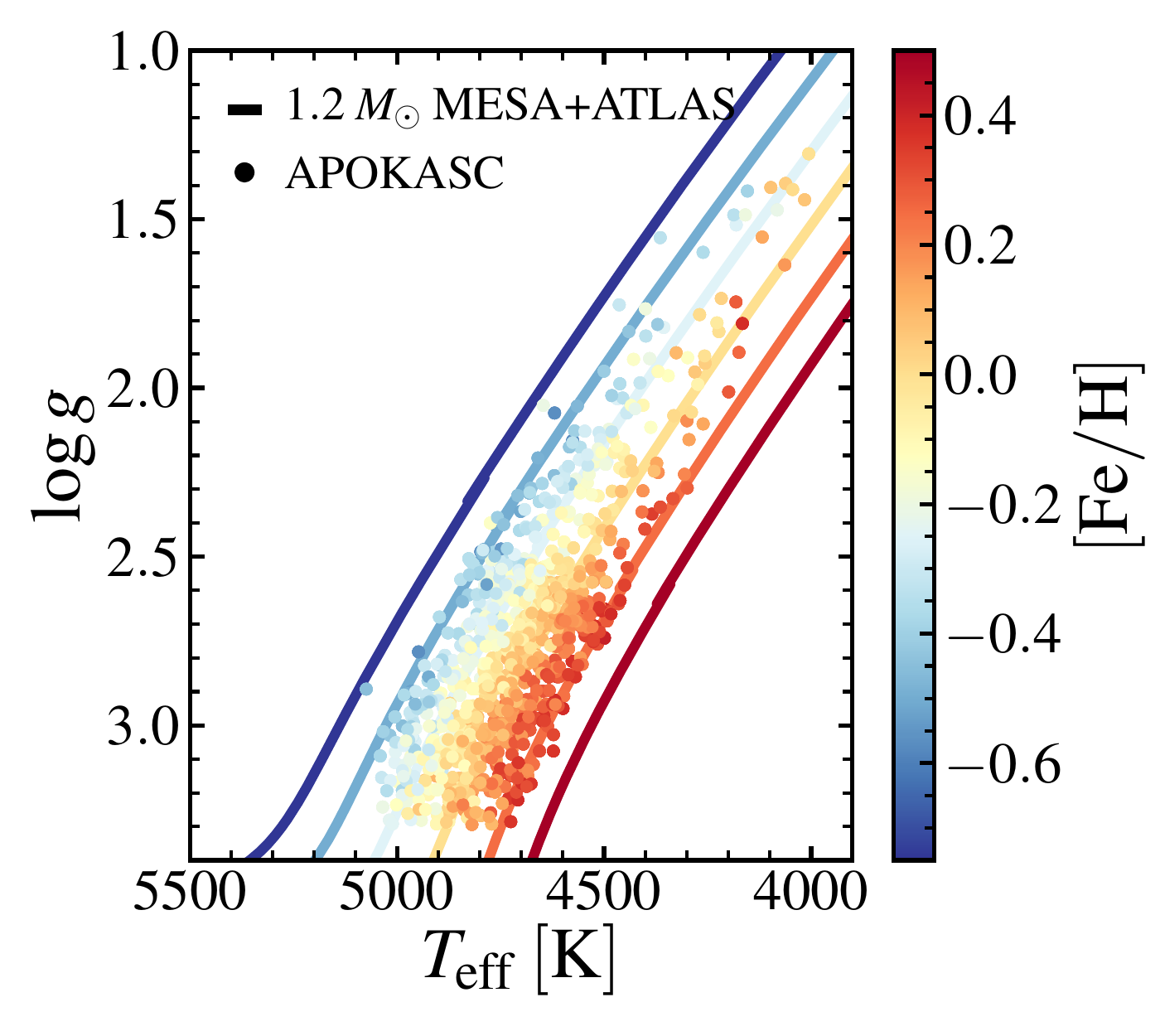}
\vspace{0.1cm}
\caption{Comparison between the observed RGB stars with masses between 1.1 and $1.3\Msun$ and \texttt{MESA} evolutionary tracks for a $1.2\Msun$ star at a range of metallicities ($\rm [Fe/H]=-0.75$ to $0.50$ in steps of $0.25$~dex). $\teff$ for the observed sample are derived from the APOGEE spectra and correspond to the post-DR13 values adopted in T17. The $\loggast$ and $M_{\rm init}$ are obtained using the asteroseismic scaling relations. The stars are selected to have roughly solar-scaled abundances ($[\rm \alpha/Fe] < 0.07$) and each star is color-coded by [Fe/H].}
\label{fig:dtobs1}
\end{figure}

Figure~\ref{fig:dtobs1} shows a comparison in the Kiel diagram between the APOKASC RGB stars and the \texttt{MESA} evolutionary tracks. We select stars that have solar-scaled abundances ($[\rm \alpha/Fe] < 0.07$) and asteroseismic masses between 1.1 and $1.3\Msun$. Each star is color-coded by its [Fe/H]. The models correspond to a $1.2\Msun$ star at a range of metallicities ($\rm [Fe/H]=-0.75$ to $0.50$ in steps of $0.25$~dex) computed with the \texttt{ATLAS} surface BC. The \texttt{MESA} RGB tracks show a moderate tension with the APOGEE $\teff$, both in terms of the overall $\teff$ shift and in the $\teff$ ``stretch'' with metallicity. This tension is quantified below.

\subsection{Effect of Boundary Conditions}
\label{s:mvd_bc}

In this section we carry out a star-by-star comparison between the APOGEE $\teff$ and the \texttt{MESA} $\teff$ along the RGB. For each star, we interpolate in [Fe/H], [$\alpha$/Fe], $\loggast$, and $M_{\rm init}$ from a grid of model evolutionary tracks to obtain $\teff$.\footnote{S18 note that [$\alpha$/Fe] and [Fe/H] reported in the T17 catalog are in fact $[\alpha$/M] and [M/H] in the in APOGEE DR13 catalog, where M is the total metallicity. The resulting error is small (0.01--0.02~dex). See Figure~7 in S18 for more details.} We perform this comparison for each of the surface BCs described in previous sections.

\begin{figure*}
\center
\includegraphics[width=0.85\textwidth]{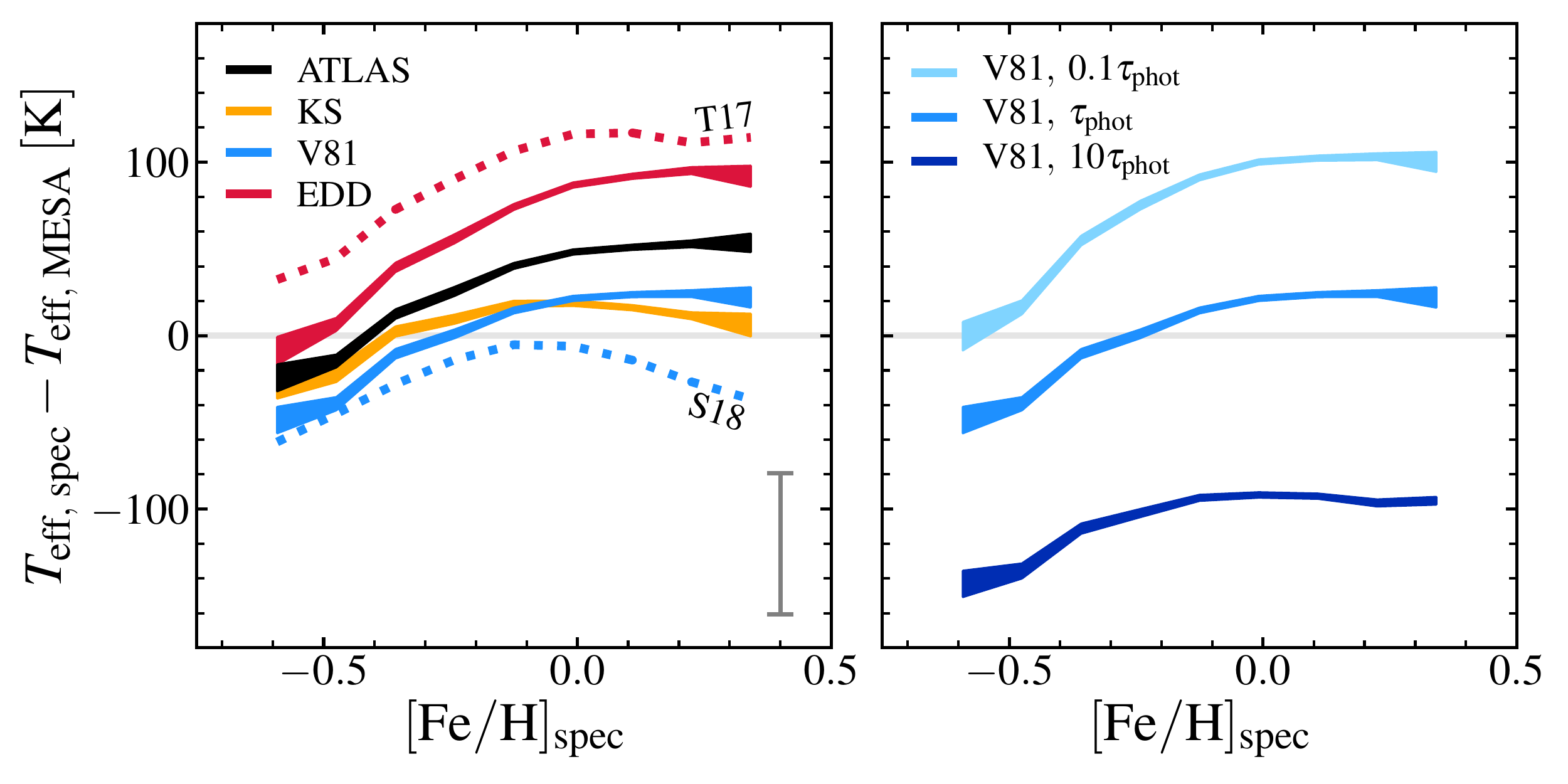}
\vspace{0.1cm}
\caption{Effect of surface boundary condition on model--observation comparisons along the RGB. Curves are mean-binned $\Delta \teff$ for the APOGEE $\teff$ values adopted in T17 and \texttt{MESA} models computed with a variety of surface boundary conditions. The line widths correspond to the error on the mean and the gray error bar represents the typical scatter within each bin. \emph{The choice of surface boundary condition and the optical depth at which it is applied in the stellar interior models induces $\approx 100~\rm K$ variation in the resulting RGB effective temperatures.} Left Panel: \texttt{MESA} models computed with different types of boundary condition. \texttt{ATLAS} is based on realistic model atmosphere computations whereas KS \citep{KrishnaSwamy1966}, V81 \citep{Vernazza1981}, and EDD \citep{Eddington1926} are analytic relations. We include the T17 and S18 $\teff$ trends in dotted lines, color-coded by their adopted model boundary condition (EDD and V81, respectively). Right Panel: Same as the left panel now showing different $\tau_{\rm base}$, the optical depth at which the surface boundary condition is applied in the stellar interior model. All three models were computed with the V81 boundary condition. Note that both the normalization and the slope change in this case.}
\label{fig:dtmobs_summary}
\end{figure*}

The resulting comparison between APOGEE and model $\teff$ is shown in Figure~\ref{fig:dtmobs_summary}. The figure shows the mean difference in $\teff$ as a function of spectroscopic metallicity. The key conclusion of this paper is that the choice of surface BC imparts a $\approx \pm100$ K systematic uncertainty in the model $\teff$ values, and this uncertainty manifests both as an overall zeropoint shift and a metallicity-dependent offset. This result amplifies and extends the conclusion in S18 that the surface BCs play a pivotal role in the comparison of models and observations.

The left panel shows $\Delta \teff$ resulting from differences in the \emph{types} of BCs employed in stellar models. The black, orange, blue, and red curves represent realistic model atmosphere BC tables from ATLAS, the KS analytic $\Ttau$ relation, the analytic fit to the V81 BC tabulation identical to what is used in S18, and the analytic Eddington Gray atmosphere used in T17, respectively. We emphasize that the differences shown here are solely due to the choice of the surface BC, and include the differences in both the $\Ttau$ relation and the solar-calibrated $\amlt$. Each curve is mean-binned, the line width corresponds to the standard error of the mean, and the gray error bar represents the typical scatter within each bin. For comparison, we include the T17 and S18 $\teff$ trends as dotted lines, color-coded by their BC of choice in model computations. Although there are $\approx 20$~K offsets due to differences in the numerous physical assumptions and input choices among \texttt{MESA}, YREC, and BaSTI, the dotted lines show similar trends compared to the corresponding BC used herein. The Eddington atmosphere results in both the largest overall offset and the strongest trend with metallicity. Notice that the KS BC results in no significant $\teff$ trend with [Fe/H], although we emphasize that there is no physical reason to prefer the KS BC over the others.

The right panel shows $\Delta \teff$ resulting from differences in $\tau_{\rm base}$, the \emph{location} (in terms of the optical depth) at which the surface BC is applied in the stellar interior model. All three models shown here were computed adopting the V81 $\Ttau$ relation, each with its own solar-calibrated $\amlt$. The light, medium, and dark blue curves represent $\tau_{\rm base}$ of $0.1\tauphot$, $\tauphot$, and $10\tauphot$, respectively. Recall that $\tauphot$ is defined to be where $T=\teff$ and represents the fiducial location of choice in most stellar evolution codes. Furthermore, since the optical depth increases with increasing depth in the stellar interior, $0.1\tauphot$ means the atmosphere model is ``grafted on'' farther out in the stellar atmosphere, and vice versa.\footnote{Unfortunately, many of the $0.1\tauphot$ models suffered from numerical convergence issues and so we adopted a simpler scheme to approximate the full grid. The light blue line in Figure~\ref{fig:dtmobs_summary} was generated by applying a $\Delta\teff$ interpolated as a function of [Fe/H] and $\logg$ at $1.2~\msun$ as estimated from Figure~\ref{fig:dtm_varyBCs_v2} to the medium blue curve ($\tauphot$). We checked the accuracy of this simple $\Delta \teff$ interpolation by comparing the $10\tauphot$ relation estimated using this method to the relation obtained from actually interpolating evolutionary tracks computed assuming $10\tauphot$ BCs. The difference is negligible, amounting to a few K for the majority of the stars.}


\section{Discussion}
\label{s:discussion}

The main result of this paper is that the choice of surface BC in solar-calibrated stellar models imparts a substantial ($\approx100$ K) uncertainty in the effective temperature distribution along the RGB. The effect of the surface BC is not a constant shift in $\teff$, but instead results in changes to $\teff$ that vary with $\logg$ and metallicity. Amongst the different models that we have explored, we believe that the model atmosphere-based BCs are the most physically realistic and hence are the most likely to be correct. The analytic $\Ttau$ relations suffer from several critical shortcomings. The Eddington model assumes gray opacities, which is well-known to be inadequate for stellar atmospheres (see also the discussion in \citealt{Chabrier1997}).\footnote{Note that gray atmospheres are adopted in asteroseismology models because grids of atmosphere tables are generally too coarsely sampled to compute eigenfunctions for pulsation modes.} The KS and V81 models, both empirical relations derived for the Sun, are meant to be employed in the radiative region, which is why they diverge strongly from the \texttt{ATLAS} models at large $\tau$ in Figure \ref{fig:Ttau}. There is no reason that scaled versions of these relations should adequately describe stars of all metallicities and $\logg$ values.

Self-consistent application of surface BCs requires that the adopted physics in the atmosphere and the interior agree in the joining region. This includes ensuring the same equation of state, sources of opacity, and treatment of convection. We note that \texttt{ATLAS} uses a somewhat different implementation of MLT \citep{Mihalas1978} compared to the one that we use in \texttt{MESA} \citep{Henyey1965}. Preliminary work (see also \citealt{Montalban2001} and \citealt{VandenBerg2008}) suggests that the attachment location of the BC does not strongly influence the resulting RGB $\teff$ as long as the treatment of convection and $\amlt$ are self-consistent between the interior and the atmosphere models. In practice, atmosphere models are computed with an internally-calibrated $\amlt$ that is not guaranteed to be consistent with that of an interior model, because this requires iterating the solar-calib rations to converge on a common value of $\amlt$ \citep{VandenBerg2008}. In future work we plan to explore a common treatment of MLT in \texttt{ATLAS} and \texttt{MESA}.

The sensitivity of the RGB $\teff$ to the location of the joining region ($0.1-10\tauphot$) in the models presented in this work, as shown in Figure \ref{fig:dtm_varyBCs_v2}, is in large part a consequence of the fact that the interiors assume grayness while most of our adopted surface BCs do not. Careful inspection of Figure \ref{fig:Ttau} reveals that the Eddington $\Ttau$ relation is most similar to the \texttt{MESA} profile extended to $\tau_{\rm base} =0.1 \tauphot$, which is a reflection of the gray assumption in both cases. For the Eddington $\Ttau$ we have found that the RGB $\teff$ is in fact {\it insensitive} to the choice of $\tau_{\rm base}$ up to the point where convection becomes important. This is not surprising --- in the limit where the adopted surface BC and the interior model obey the same $\Ttau$ relation, the resulting model $\teff$ (and $L$, $R$, etc.) should be insensitive to the location of the joining region.

These considerations lead us to suggest that the joining region between the atmosphere and the interior should be placed deeper in the atmosphere than is commonly adopted, ideally where the gray opacity assumption is valid and the atmosphere and interior treatments of the opacity and convection are consistent (see \citealt{Chabrier1997} who advocate this approach for modeling cool dwarfs).\footnote{Note that $\teff$ reported by \texttt{MESA} is evaluated using the Stefan-Boltzmann equation along with $L$ and $R$ at $\tau_{\rm base}$, the outermost point in the interior model. We do not solve for the temperature at $\tauphot$ when $\tau_{\rm base} \neq \tauphot$. The resulting error is quite small and, unsurprisingly, depends on the surface gravity. For a $1~\msun$ star at the main sequence turn off, this introduces an error of $<1$~K. The error is slightly larger for an RGB star ($\sim5$~K), but this is still a $0.1\%$ effect.} At sufficiently high values of $\tau$, atmosphere models are convective and follow a simple adiabat, and are thus easy to match seamlessly to an adiabat in the interior model. While there is a physically-motivated preference for $\tau > \tau_{\rm phot}$ to ensure the validity of the diffusion approximation, the surface BC also cannot be applied arbitrarily deep. In the case of analytic $\Ttau$ relations considered in this work, their attachment location should not exceed the depth of the onset of convection since they do not model the effects of convection. Moreover, there is the more fundamental concern that a simple solar-scaled or purely analytic description of the stellar atmosphere is woefully insufficient for accurately representing the interior. Similarly, current generation of model atmospheres carry out computationally intensive radiative transfer calculations by simplifying other physics, e.g., assuming an ideal gas, and thus should not replace the detailed interior modeling in significant portions of the outer layers. To summarize, while we propose applying the surface BC deeper than at the photosphere and strongly favor the use of atmosphere models, there is still some ambiguity associated with the appropriate attachment location of the BC. We leave for future work a detailed investigation of this issue.

We now return to the question that originally prompted this investigation --- whether or not there is evidence for variation in $\amlt$ with metallicity amongst first ascent giants. T17 found a metallicity-dependent $\teff$ discrepancy when comparing data to models (red dotted line in Figure~\ref{fig:dtmobs_summary}), which they interpreted as evidence for a metallicity-dependent $\amlt$. Our \texttt{MESA} models show a wide range of behaviors depending on the adopted surface BC, and recover the results of both T17 and S18 as extremes of the theoretical range. The KS $\Ttau$ relation shows nearly perfect agreement with the APOGEE data while the Eddington $\Ttau$ relation produces a trend of $\approx100$ K dex$^{-1}$. Other models show behavior in between these two limits. More worrisome from our point of view is the result that varying $\tau_{\rm base}$ for a fixed $\Ttau$ relation results in both a different normalization {\it and} slope in the metallicity-dependent $\Delta\teff$ comparison. Given that the choice of surface BCs imparts a systematic uncertainty at least as large as the model-data tension reported in T17, we conclude that it is premature to appeal to variation of $\amlt$ with metallicity. Indeed, any conclusion that requires model RGB $\teff$ to be more accurate than $\approx100$ K must await a thorough investigation of the proper treatment of surface BCs in stellar evolution models. Moreover, these results demonstrate that solar-calibration alone is insufficient to guarantee accurate models in other parts of the Hertzsprung-Russell diagram, underscoring the need for other ``ground truth'' observational constraints.
\newline


\section{Summary}
\label{s:summary}

In this work we explored the impact of the adopted surface boundary condition on the $\teff$ of red giant stars. We computed 1D stellar evolutionary tracks using \texttt{MESA} with several implementations of surface BCs, including three analytic relations commonly used in the literature (\citealt{Eddington1926} ``gray'', \citealt{KrishnaSwamy1966}, and \citealt{Vernazza1981}) and one set of realistic model atmosphere BC tables from \texttt{ATLAS} \citep{Kurucz1970,Kurucz1993}. We performed solar-calibration for each type of the surface boundary condition to obtain an appropriate $\amlt$. Even though all models line up perfectly around the solar value---by construction---both the type of boundary condition {\it and} the location at which it is applied to the 1D stellar interior model yield $\approx \pm$100~K, metallicity- and $\logg$-dependent changes to the $\teff$ distribution along the RGB. This clearly demonstrates that solar-calibration alone is an insufficient check on the accuracy of the stellar models. In light of these results, we caution against attempting to interpret any data-model $\teff$ discrepancies at the $\approx \pm$100~K level or less until the ambiguity in the surface boundary condition presented herein is resolved.


\acknowledgments 
We thank the anonymous referee, Santi Cassisi, Lars Bildsten, and Jamie Tayar for useful discussions and insightful comments on this manuscript. We would like to thank Bill Paxton and the \texttt{MESA} community for making this work possible. We also wish to acknowledge and thank the Kavli Institute for Theoretical Physics at UC Santa Barbara, where most of this work took place. This research was supported in part by the National Science Foundation under Grant No. NSF PHY-1748958. This paper is based upon work supported by the National Aeronautics and Space Administration (NASA) under Contract No. NNG16PJ26C issued through the WFIRST Science Investigation Teams Program.

\software{MESA \citep{Paxton2011, Paxton2013, Paxton2015,Paxton2018}; matplotlib \citep{Hunter2007_matplotlib}; astropy \citep{Astropy2013}; scipy \citep{Jones2001}}

\bibliography{latest_bibtex}

\end{document}